\documentclass{ws-ijmpa}

\begin{document}

\markboth{R. Garc\'\i{}a-Mart\'\i{}n {\em et al.}}
{Precise dispersive data analysis of the $f_0(600)$ pole}

%
\catchline{}{}{}{}{}
%

\title{Precise dispersive data analysis of the $f_0(600)$ pole}

\author{\footnotesize R. Garc\'\i{}a-Mart\'\i{}n$^{\star}$$^,$\footnote{E-mail address: 
rgmar@fis.ucm.es}~, 
R.~Kami\'nski$^{\%}$, 
J.~R.~Pelaez$^{\star}$ 
}

\address{
$^{\star}$Dpto. F\'\i{}sica Te\'orica II, Universidad Complutense, Madrid, Spain\\ 
$^{\%}$Institute of Nuclear Physics, Polish Academy of Science, Cracow, Poland\\
}

\maketitle


\begin{abstract}
We review how the use of recent precise data on kaon decays together with
forward dispersion relations (FDR) and Roy's equations allow us to determine the
sigma resonance pole position very precisely, by using only experimental
input. In addition, we present preliminary results for a modified set of Roy-like
equations with only one subtraction, that show a remarkable improvement in the precision
around the $\sigma$ region.
We also improve the matching between the parametrizations at
low and intermediate energy of the $S0$ wave, and show that the effect of this
on the sigma pole position is negligible.

\end{abstract}

\section{Introduction}

The values quoted in the Particle Data Table for the sigma or $f_0(600)$ resonance
mass and width,
based on both pole position and Breit-Wigner parameter determinations
are very widely spread, with an estimated mass and half width of~\cite{PDT2006}:
\begin{equation}
  \sqrt{s_\sigma}\equiv M_\sigma -i\, \Gamma_\sigma/2
\simeq (400 - 1200) -i (250 - 500)\mbox{ (MeV)}.
\end{equation}
This large uncertainty is mainly due to the fact that old 
data sets for pion-pion scattering are poor and often contradictory.
Moreover, the choice of data sets varies among different works.
To make things worse, there is quite a variety of different ways to
extrapolate the data on the real axis to the complex plane, and the
pole position of the sigma is greatly affected by model dependences.

This said, model independent techniques for extrapolating
amplitudes from the real axis onto the complex plane exist in the form of
dispersion relations, which allow us to analytically continue an
amplitude away from the real axis provided we know its imaginary
part for physical values of the energy. These dispersive techniques
have already been
successfully used for predicting the position of the sigma pole,
with a remarkable agreement among the different works:
\begin{eqnarray}
  & 440 - i\,245~\mbox{MeV}&\mbox{Dobado, Pelaez (1997)~\cite{Dobado:1996ps}}\\
  & 470\pm50 - i\,260\pm25~\mbox{MeV}&\mbox{Zhou {\em et al.} (2005)~\cite{Zhou:2004ms}}
\end{eqnarray}

In particular, there exists a dispersive representation that incorporates crossing
exactly,
written by Roy~\cite{Roy:1971tc}, which involves only the partial wave amplitudes.
Roy's equations have already been used to predict the position of the sigma pole
from the theoretical predictions of ChPT~\cite{Caprini:2005zr}, obtaining:
\begin{equation}
  \sqrt{s_\sigma}= 441_{-8}^{+16}-i\,272_{-19.5}^{+9}\mbox{ MeV}
\end{equation}

In addition, the data coming from the E865 collaboration at Brookhaven~\cite{Pislak:2001bf},
and especially the recently published data from NA48/2~\cite{Batley:2007zz} provide
us with very precise data on pion-pion scattering at very low energies,
These allow us to obtain very reliable parametrizations of the S0 wave at low
energy~\cite{PelaezTalk}, from which the scattering lengths can be
directly extracted~\cite{Yndurain:2007qm} with a remarkable precision and
in good agreement with the theoretical predictions of
ChPT~\cite{Caprini:2005zr}.

Our aim is thus to perform a dispersive analysis, including all available
experimental data, in order to give a precise and model independent
determination of the sigma pole position, by using exclusively data,
analyticity and crossing symmetry. We use both Forward Dispersion
Relations (FDR) and Roy's equations, without assuming ChPT,
so that we can actually test its predictions.

\section{Approach and results}

The details on the parametrizations used for the data have been explained fully
in Ref.~\cite{Kaminski:2008qe}, that we will denote
by KPY08. It is enough to say here that two different
sets of parameters are considered:
\begin{itemize}
\item {\em Unconstrained Fits to Data} (UFD),
in which each partial wave is fitted independently.
This set satisfies both FDR and Roy's equations within the experimental
errors in all waves except the Roy equation for the S2 wave, 
for which the deviation is
about $1.3\,\sigma$, and 
the antisymmetric FDR above 930 MeV by a couple of standard deviations.
\item{\em Constrained Fits to Data} (CFD), obtained
 by constraining the fits to satisfy simultaneously FDR and Roy's
equations, so that all waves are correlated. The CFD set provides
a remarkably precise and reliable description of the experimental data,
and at the same time satisfy the analytic properties remarkably well.
\end{itemize}
These two sets provide a reliable parametrization for the imaginary
part of the partial waves that we need as input for Roy's equations.

An elastic resonance has an associated pole on the second Riemann sheet of the
complex plane S-matrix, which, as it is well known, corresponds by unitarity
to a zero on the first sheet. As usual then, we just need to look numerically for
zeroes of the S-matrix on the physical sheet,
$S_0^0(s)=1+2i\sigma(s) t_0^0(s)$,
where the analytic extension
of the partial wave amplitudes away from the real axis is given by
Roy's equations,
whose domain of validity has been shown to cover the
region of the complex plane where the sigma lies~\cite{Caprini:2005zr}.

Taking the UFD set as the input for Roy's equations, we find an S-matrix
zero at $\sqrt{s}=(426\pm 25)-i(241 \pm 17)$~MeV. However, Roy's equations
are not completely satisfied by this data set, thus the pole position will
be much more reliable if the input satisfies the equations,
as it is the case for the CFD set. In this case we find:
\begin{equation}
  \label{roy-cfd-pole}
  \sqrt{s_\sigma} = (456 \pm 36) - i (256 \pm 17)\mbox{ MeV},
\end{equation}
which still has big uncertainties due to the strong dependence of Roy's
equations on the scattering lengths, in particular of the $a_0^2$, which
is known with less precision. These values are, however, subject to further
improvement and should be considered preliminary. It should also be noted
that they are in perfect agreement with the theoretical prediction
by Caprini {\em et al.} of
$\sqrt{s_\sigma}=441_{-8}^{+16}-i\,272_{-19.5}^{+9}$.

\section{Work in progress}
The three authors of this work together with F.~J.~Yndur\'ain (see, i.e.,
\cite{Kaminski:2008fu} in this conference or Ref.~\cite{PelaezTalk}) have derived
a modified set of Roy-like equations which are based on once-subtracted
dispersion relations -- Roy's equations are twice subtracted.
The motivation for these new equations is that
their uncertainties are smaller than
for standard Roy's equations in the region above $\sim 400$~MeV,
which is of interest for our work.
This allows us to obtain the position of
the sigma pole from Constrained Fits to Data
with higher accuracy than by constraining with the standard Roy's equations
alone. In addition, they allow us to better describe the $f_0(980)$ region,
as the errors there are now much smaller and some parametrizations could
be now discarded.

\begin{figure}[t]
  \begin{center}
    \includegraphics[width=7cm,height=10cm,angle=-90]{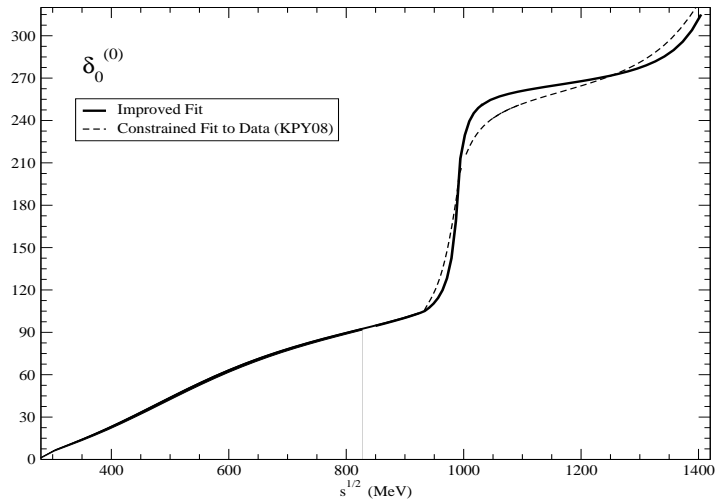}
    \caption{S0 wave phase shifts. We show both the previous fit in KPY08
and a new preliminary fit that improves the matching between the conformal
and K-matrix regions.}
    \label{fig-delta00}
  \end{center}
\end{figure}

We have already performed~\cite{PelaezTalk} a preliminary 
 Constrained Fit to Data (CFD-II) in which these new
equations are also imposed as new constraints within errors.
Moreover, following a suggestion in~\cite{Leutwyler}
we have improved the matching of the low and intermediate ($f_0(980)$)
regions at $\sqrt{s}=932$~MeV by imposing continuity not only on the phase
shifts, but also on the derivative.
This gives rise to a new set of parameters which better encode the experimental
information together with unitarity, analyticity and crossing symmetry,
therefore allowing us to obtain a more precise and reliable determination
of the sigma pole, which for this preliminary CFD-II set is:
\begin{eqnarray}
  \label{eq:sigmapole}
&&  \sqrt{s_\sigma} = (459 \pm 47) - i (257 \pm 18)\mbox{ MeV}, \;  \hbox{\rm (preliminary from Roy Eqs.)}
\\
&&  \sqrt{s_\sigma} = (461 \pm 14) - i (255 \pm 14)\mbox{ MeV}, \;  \hbox{\rm (preliminary from GKPY)}
\label{sigmagkpy}
\end{eqnarray}
to be compared with the result from the GKPY equations of
$\sqrt{s_\sigma} = (461 \pm 13) - i (254 \pm 14)$~MeV, obtained with
the CFD data set in which the improved matching condition of a continuous first
derivative was not imposed.
We can see that, although the phase shift does change above 932~MeV
(see Fig.~\ref{fig-delta00}),
the pole position is almost unaffected.
A full analysis including these new equations should be complete within the next few months.

\section{Aknowledgements}
This work is dedicated to the memory of Prof. F. J. Yndur\'ain.
We thank H. Leutwyler for his comment about the matching condition, and
the organizers for creating the nice scientific atmosphere of the event
and the Spanish research contracts PR27/05-13955-BSCH, 
FPA2004-02602, UCM-CAM 910309 and BFM2003-00856 for partial financial support.


\begin{thebibliography}{9}

\expandafter\ifx\csname natexlab\endcsname\relax\def\natexlab#1{#1}\fi
\providecommand{\enquote}[1]{``#1''}
\expandafter\ifx\csname url\endcsname\relax
  \def\url#1{\texttt{#1}}\fi
\expandafter\ifx\csname urlprefix\endcsname\relax\def\urlprefix{URL }\fi
\providecommand{\eprint}[2][]{\url{#2}}

\bibitem{PDT2006}
W.~M. Yao, et~al., \emph{J. Phys.} \textbf{G33}, 1--1232 (2006).

\bibitem{Dobado:1996ps}
A.~Dobado, and J.~R. Pelaez, \emph{Phys. Rev.} \textbf{D56}, 3057--3073 (1997).

\bibitem{Zhou:2004ms}
Z.~Y. Zhou, et~al., \emph{JHEP} \textbf{02}, 043 (2005).

\bibitem{Roy:1971tc}
S.~M. Roy, \emph{Phys. Lett.} \textbf{B36}, 353 (1971).

\bibitem{Caprini:2005zr}
I.~Caprini, G.~Colangelo, and H.~Leutwyler, \emph{Phys. Rev. Lett.}
  \textbf{96}, 132001 (2006).

\bibitem{Pislak:2001bf}
S.~Pislak, et~al., \emph{Phys. Rev. Lett.} \textbf{87}, 221801 (2001).

\bibitem{Batley:2007zz}
J.~R. Batley, et~al.  (2008), cERN-PH-EP-2007-035.

\bibitem{Kaminski:2008fu}
  R.~Kaminski, R.~Garcia-Martin, P.~Grynkiewicz, J.~R.~Pelaez and F.~J.~Yndurain,
  arXiv:0809.4766 [hep-ph].

\bibitem{PelaezTalk}
J.~R.~Pelaez {\em et al.},
  AIP Conf.\ Proc.\  {\bf 1030} (2008) 257

\bibitem{Yndurain:2007qm}
F.~J. Yndurain, R.~Garcia-Martin, and J.~R. Pelaez, \emph{Phys. Rev.}
  \textbf{D76}, 074034 (2007).

\bibitem{Kaminski:2008qe}
  R.~Kaminski, J.~R.~Pelaez and F.~J.~Yndurain,
  Phys.\ Rev.\  D {\bf 77} (2008) 054015.

\bibitem{Leutwyler}
  H.~Leutwyler,
  AIP Conf.\ Proc.\  {\bf 892} (2007) 58.
\end{thebibliography}
\end{document}